\documentstyle[12pt]{article}
\begin{document}

\author{Marc Henneaux \\
Facult\'e des Sciences, Universit\'e Libre de Bruxelles,\\
B-1050 Bruxelles, Belgium\\
and \\
Centro de Estudios Cient\'\i ficos de Santiago,\\
Casilla 16443, Santiago 9, Chile}
\title{Uniqueness of the Freedman-Townsend Interaction Vertex For Two-Form
Gauge
Fields}
\date{}
\maketitle

\begin{abstract}
Let $B_{\mu \nu }^a$ ($a=1,...,N$) be a system of $N$ free two-form gauge
fields, with field strengths $H_{\mu \nu \rho }^a = 3 \partial _{[\mu
}B_{\nu \rho ]}^a$ and free action $S_0$ equal to $(-1/12)\int d^nx\
g_{ab}H_{\mu \nu \rho }^aH^{b\mu \nu \rho }$ ($n\geq 4$). It is shown that
in $n>4$ dimensions, the only consistent local interactions that can be
added to the free action are given by functions of the field strength
components and their derivatives (and the Chern-Simons forms in $5$ mod $3$
dimensions). These interactions do not modify the gauge invariance $B_{\mu
\nu }^a\rightarrow B_{\mu \nu }^a+\partial _{[\mu }\Lambda _{\nu ]}$ of the
free theory. By contrast, there exist in $n=4$ dimensions consistent
interactions that deform the gauge symmetry of the free theory in a non
trivial way. These consistent interactions are uniquely given by the
well-known Freedman-Townsend vertex. The method of proof uses the
cohomological techniques developed recently in the Yang-Mills context to
establish theorems on the structure of renormalized gauge-invariant
operators.
\end{abstract}

\newpage\

Two-form gauge fields play a central r\^{o}le in string theory and in
various supergravity models. They provide generalizations of Maxwell theory
that are not only interesting in their own right, but which also shed a new
light on the properties and the structure of the Maxwell theory itself.

In the absence of interactions, two-form gauge fields are described by the
action
\begin{equation}
S_0=-\frac 1{2\cdot 3!}\int d^nx\ g_{ab}H_{\mu \nu \rho }^aH^{b\mu \nu \rho }
\label{freeaction}
\end{equation}
where $g_{ab}$ is an invertible, symmetric matrix and where $H_{\mu \nu \rho
}^a$ are the field strength components,
\begin{equation}
H_{\mu \nu \rho }^a=\partial _\mu B_{\nu \rho }^a+\partial _\nu B_{\rho \mu
}^a+\partial _\rho B_{\mu \nu }^a.  \label{fieldstrength}
\end{equation}
The free theory is invariant under the abelian gauge transformations
\begin{equation}
\delta _\Lambda B_{\mu \nu }^a=\partial _\mu \Lambda _\nu ^a-\partial _\nu
\Lambda _\mu ^a,  \label{gaugeinv}
\end{equation}
while the free field equations of motion are
\begin{equation}
\frac{\delta S_0}{\delta B_{\mu \nu }^a}=0\Leftrightarrow \partial _\rho
H^{a\rho \mu \nu }=0.  \label{equmotion}
\end{equation}
The gauge symmetries (\ref{gaugeinv}) are reducible, since if one takes
\begin{equation}
\Lambda _\nu ^a=\partial _\nu \varepsilon ^a  \label{reducibility}
\end{equation}
in (\ref{gaugeinv}), one gets a vanishing variation for $B_{\mu \nu }^a$.

This paper investigates the possible consistent couplings that can be
introduced among a system of $N$ free two-forms $B_{\mu \nu }^a$, without
introducing further dynamical fields\footnote{%
New types of couplings not discussed here are well known to be allowed if
there are further fields besides the $N$ two-forms.}$^{,}$\footnote{%
An interaction term $\Delta S_0$ is said to be consistent if one can modify
the gauge symmetries and the reducibility identities in such a way that the
modified action $S_0+\Delta S_0$ is invariant under the modified gauge
symmetries, which are required to fulfill the same number of (possibly
modified) reducibility identities. We shall consider here only consistent
interactions that can be obtained by continuous deformation of the action.},
for all spacetime dimensions $n\geq 4$ (the case $n<4$, for which the
two-forms $B_{\mu \nu }^a$ carry no degree of freedom, is not considered).
Our main result is that the abelian gauge transformations (\ref{gaugeinv})
are extremely rigid, in the sense that no consistent interaction can deform
them. The sole exception occurs in four dimensions, where the
Freedman-Townsend interaction vertex \cite{FT} defines the only consistent
interaction that deforms non trivially the gauge transformations of a set of
$N$ free two-forms. [Note, however, that even in that case, the algebra of
the deformed gauge transformations remains abelian]. This situation is in
sharp contrast to what happens with one-form gauge fields, for which
consistent deformations of the gauge symmetries - namely, Yang-Mills
deformations - exist in any number of spacetime dimensions.

The difference between one-forms and two-forms, and, more generally, between
one-forms and $p$-forms, was pointed out already some time ago in \cite{CT}
from a geometrical viewpoint. The argument goes as follows. One may regard
two-form gauge fields as connections in the space of, say, closed strings. A
path in loop space defines a two-dimensional surface in spacetime whose
boundaries are the initial and final string configurations. Now, the same
spacetime surface can be swept out differently by the string and corresponds
therefore to many different paths in loop space. If one requires the
parallel transport from the initial to the final string configurations to
depend only on the spacetime surface swept out by the string, and not on the
actual details of how it is swept out, then, one is inevitably led to the
conclusion that the symmetry group must be abelian, with gauge
transformations (\ref{gaugeinv}) for the two-form connection.

One may object to this geometrical argument that there is no reason to
assume that the two-forms, which are local fields in spacetime, should define
a  connection in loop space. Actually, the Freedman-Townsend model evades it
precisely because in it, the two-form gauge fields have no immediate
connection interpretation. However, this is the only exception and one can
thus conclude that the geometrical argument is quite powerful.

In order to establish the rigidity of the gauge transformations (\ref
{gaugeinv}) in $n\geq 5$ dimensions and the uniqueness of the
Freedman-Townsend deformation of the gauge symmetry in four dimensions, I
shall follow the antifield-BRST approach to the analysis of consistent
deformations developed in \cite{BH1}. This approach is based on the
observation that a consistent deformation of the free action $S_0$ and of
its gauge symmetries defines a deformation of the corresponding solution of
the master equation that preserves both the master equation and the
field-antifield spectrum. Thus, if $S_0+g\int d^nx\ a_0+O(g^2)$ defines a
consistent deformation of $S_0$, with deformed gauge symmetry $\delta
_\Lambda ^{\prime }B_{\mu \nu }^a=\partial _\mu \Lambda _\nu ^a-\partial
_\nu \Lambda _\mu ^a+g\Delta _{\mu \nu }^a+O(g^2)$, then one has for the
solution of the master equation
\begin{equation}
S\rightarrow S+g\int d^nx\ a+O(g^2),
\end{equation}
with
\begin{equation}
\left( S,S\right) =0,\ \left( S+g\int d^nx\ a+O(g^2),S+g\int d^nx\
a+O(g^2)\right) =0.  \label{master}
\end{equation}
Here, $S$ and $a$ are respectively given by
\begin{equation}
S=S_0-\int d^nx\left( 2\partial _\mu C_\nu ^aB_a^{*\mu \nu }+\partial _\mu
\eta ^aC_a^{*\mu }\right)
\end{equation}
and
\begin{equation}
a=a_0-\widetilde{\Delta }_{\mu \nu }^aB_a^{*\mu \nu }+\hbox{``more"}.
\label{expansion}
\end{equation}
In these formulas, the $C_\mu ^a$ are the ghosts and have ghost number one.
The $\eta ^a$ are the ghosts of ghosts of ghost number two, which appear
because the gauge transformations are reducible \cite{Townsend,Siegel}. The $%
B_a^{*\mu \nu }$, $C_a^{*\mu }$ and $\eta _a^{*}$ are the corresponding
antifields and have respective ghost numbers $-1$, $-2$ and $-3$ \cite{BV,HT}%
. The ghost number itself can be viewed as the difference between the pure
ghost number and the antighost number, where the non trivial assignments
are:
\begin{eqnarray}
\hbox{pure gh}(C_\mu ^a) &=&1,\;\hbox{pure gh}(\eta ^a)=2, \\
\hbox{antigh}(B_a^{*\mu \nu }) &=&1,\;\hbox{antigh}(C_a^{*\mu })=2,\;%
\hbox{antigh}(\eta _a^{*})=3.
\end{eqnarray}
The term $\widetilde{\Delta }_{\mu \nu }^a$ in (\ref{expansion}) is obtained
by replacing the gauge parameter $\varepsilon _\mu ^a$ by the ghosts $C_\mu
^a$ in $\Delta _{\mu \nu }^a$. Moreover, ``more'' stands for terms of
antighost number $\geq 2$.

The deformed solution of the master equation contains information not only
about the interaction term $a_0$ added to the free action, but also about
the new form of the gauge symmetries, their algebra, and their reducibility
identities. More precisely, in the expansion (\ref{expansion}), the first
antifield-dependent terms have the following interpretation: the
coefficients of the terms in $B^{*}C$ define the deformation of the gauge
symmetry; the coefficients of the terms in $C^{*}CC$ correspond to the
deformation of the gauge algebra; the coefficients of the terms in $%
C^{*}\eta $ define the deformation of the reducibility coefficients; the
reducibility of the new gauge symmetry holds only on-shell if one of the
terms in  $B^{*}B^{*}\eta $ does not vanish etc (all this to order $g$).
Thus, a consistent deformation modifies neither the gauge transformations
nor their algebra or reducibility relations, if  the term $g\int d^nx\ a$
added to the solution of the master equation for the free theory does not
contain the antifields and reduces to $g \int d^n x a_0$.

The interest of reformulating the problem of consistent interactions through
the master equation is that the consistency of the interaction imposes
equations on the terms added to $S$ that have a direct cohomological
interpretation.\ \ The problem  can then be investigated by the powerful
machinery of homological algebra. Indeed, it is easy to see that equation (%
\ref{master}) holds to order $g$ if and only if the first order deformation $%
a$ is a BRST-cocycle modulo $d$, i.e., fulfills

\begin{equation}
sa=\partial _\mu k^\mu ,  \label{BRSTcocyc}
\end{equation}
for some $k^\mu $. Furthermore, if $a$ is a coboundary modulo $d$, $%
a=sb+\partial _\mu m^\mu $ for some $b$ and $m^\mu $, then the deformation
is trivial in the sense that it can be eliminated by field redefinitions.
Accordingly, non-trivial first-order consistent interactions are uniquely
determined by the elements of the cohomological group $H^0(s|d)$, and
vice-versa. Here, the BRST differential $s=\delta +\gamma =(S,.)$ is the
BRST differential of the free theory acting in the space of fields and
antifields.\ \ One has $s=\delta +\gamma $ with
\begin{eqnarray}
\delta B_{\mu \nu }^a &=&0,\quad \gamma B_{\mu \nu }^a=\partial _\mu C_\nu
^a-\partial _\nu C_\mu ^a, \\
\delta C_\mu ^a &=&0,\quad \gamma C_\mu ^a=\partial _\mu \eta ^a, \\
\delta \eta ^a &=&0,\quad \gamma \eta ^a=0
\end{eqnarray}
and
\begin{eqnarray}
\delta B_a^{*\mu \nu } &=&\frac 12g_{ab}\partial _\rho H^{b\rho \mu \nu
},\quad \gamma B_a^{*\mu \nu }=0, \\
\delta C_a^{*\mu } &=&2\partial _\nu B_a^{*\mu \nu },\quad \gamma C_a^{*\mu
}=0 \\
\delta \eta _a^{*} &=&\partial _\mu C_a^{*\mu },\quad \gamma \eta _a^{*}=0.
\end{eqnarray}
According to the general antifield theory,  the BRST differential has been
split as the sum of the Koszul-Tate differential $\delta $ and the
longitudinal exterior derivative $\gamma $ \cite{HT}. Taking $a$ to be a
BRST-cocycle modulo $d$ makes $S+g\int d^nx\ a$ a solution of the master
equation up to order $g$ included.

There are further conditions on $a$ which express that  $S+g\int d^nx\ a$
can be completed by terms of higher order so as to be a solution of the
master equation to all orders (``the first-order deformation $g\int d^nx\ a$
is not obstructed at higher orders'') .\ \ The condition at order $g^2$ is
that the antibracket $(\int d^nx\ a,\int d^nx\ a)$ should be exact in the
space of local functionals \cite{BH1}. This condition is automatically
fulfilled if $a$ does not contain the antifields (and so, reduces to $a_0$)
since then the antibracket $(\int d^nx\ a,\int d^nx\ a)$ vanishes.\ \ When
this happens,  $S+g\int d^nx\ a$ is a solution of the master equation to all
orders, without further higher order corrections.

Since the interaction $a_0$ deforms the gauge transformations if and only if
the corresponding BRST-cocycle $a$ modulo $d$ contains the antifields%
\footnote{%
If there is no modification of the gauge transformations, the deformation
does not contain the antifields $B_a^{*\mu \nu }$.\ \ Then, the reducibility
identities may also be assumed to be unchanged, so that the deformation does
not contain any antifield at all.}, the central question is whether one can
eliminate the antifields from any solution of (\ref{BRSTcocyc}) by adding a
BRST-coboundary modulo $d$. If that can be done, then there is no
deformation of the gauge transformations when one adds the interaction $%
g\int d^nx\ a=g\int d^nx\ a_0$ to the free action.

In order to examine whether the antifields can be eliminated from a
BRST-cocycle modulo $d$, one can follow the same method as in \cite{BBH1},
where the antifield-dependence in the Yang-Mills case was shown to be
controlled by the cohomological groups $H_j(\delta |d)$ of the $\delta $ mod
$d$ cohomology.

So, let $a$ be a solution of the cocycle condition (\ref{BRSTcocyc}), which
can be expanded according to the antighost number,

\begin{equation}
a_0\rightarrow a=a_0+a_1+a_2+...+a_k,\quad \hbox{antigh}(a_i)=i.
\label{solu}
\end{equation}
It is easy to prove, just as in the Yang-Mills case, that the last term $a_k$
can be assumed to be annihilated by $\gamma $ and to depend only on the
curvature components $H_{\mu \nu \rho }^a$, the antifields, their
derivatives, and the undifferentiated ghost $\eta ^a$ of ghost number two.
Thus, since antigh$(a_k)=k$ and gh$(a_k)=0$, the pure ghost number of $a_k$
is equal to $k$, which forces $k$ to be even ($a_k$ involves only the ghosts
$\eta ^a$). Assume $k=2l>0$, $a_k=\mu _{b_1b_2...b_l}\ \eta ^{b_1}\eta
^{b_2}...\eta ^{b_l}$. Now, the coefficients $\mu _{b_1b_2...b_l}$ are
easily verified to define elements of $H_k(\delta |d)$ ($\delta \mu
_{b_1b_2...b_l}+\partial _\mu \nu _{b_1b_2...b_l}^\mu =0$).  Again like
in the Yang-Mills case, the last term $a_k$ in $a$ can be removed from $a$
by an allowed redefinition, $a\rightarrow a+sb+\partial _\mu c^\mu $, iff
the $\mu _{b_1b_2...b_l}$ are trivial in the invariant cohomology $%
H_k(\delta |d)$, i.e., of the form $\delta \sigma _{b_1b_2...b_l}+\partial
_\mu \rho _{b_1b_2...b_l}^\mu $, where $\sigma _{b_1b_2...b_l}$ and $\rho
_{b_1b_2...b_l}^\mu $ involve, like $\mu _{b_1b_2...b_l}$, only the
curvature components, the antifields and their derivatives.

The crucial theorem upon which our result relies is:

\vspace{3pt}

\noindent
{\bf Theorem 1 :} {\it In spacetime dimensions }$n\geq 5${\it , the
cohomogical groups }$H_k(\delta |d)$ {\it vanish for }$k\neq 0${\it ,
}$k\neq 1$ {\it and }$k\neq 3$ {\it in the space of functions of the
curvature components, the antifields and their derivatives,}
\begin{equation}
H_k(\delta |d)=0,\ k\neq 0,\ k\neq 1,\ k\neq 3\quad (n\geq 5)
\end{equation}

\vspace{3pt}

\noindent
{\bf Proof : }The proof of this theorem is given for $k>3$ in \cite{BBH2}%
. The case $k=2$ is then easily taken care of by using the isomorphisms $%
H_0^1(d|\delta )\simeq H_1^2(\delta |d)\simeq ...\simeq H_4^5(\delta
|d)\simeq 0$, also proven in \cite{BBH2}. \ \ The details will be given in
\cite{HKS1}.

\vspace{3pt}

It follows from this theorem that one can remove $a_k$ unless $k=1$ or $k=3$%
. But we have seen that $k$ must be even.\ \ Thus the possibilities $k=1$
and $k=3$ do not arise, and the obstructions to removing the antifields from
$a$ are not encountered. This establishes:

\vspace{3pt}

\noindent
{\bf Theorem 2 :} {\it At ghost number zero and in }$n\geq 5$ {\it
dimensions, the most general solution of the cocycle condition (\ref
{BRSTcocyc}) does not depend on the antifields (up to trivial terms). Thus,
up to redefinitions, the most general consistent deformation of the free
action (\ref{freeaction}) is gauge invariant under the gauge transformations
(\ref{gaugeinv}) of the free theory (}$\gamma \int d^nx\ a_0=0${\it ).
The deformed action }$S_0+g\int d^nx\ a_0$ {\it  is consistent without
need for terms of order }$g^2$ {\it or higher. The interacting theory is
invariant under the same gauge transformations as the free theory.}

\vspace{3pt}

One can say more about the solutions of $\gamma \int d^nx\ a_0=0$, i.e. $%
\gamma a_0=\partial _\mu m^\mu $ for some $m^\mu $ \cite{HKS2}.\ \ These
fall into two classes, like in the Yang-Mills case.\ \ First, there are the
solutions for which $m^\mu $ vanishes (or can be made to vanish by
redefinitions), $\gamma a_0=0$.\ \ These are just the polynomials in the
field strength components $H_{\mu \nu \rho }^a$ and their derivatives.
Second, there are the solutions with a non trivial $m^\mu $.\ \ These are
given, up to solutions of the previous type, by the Chern-Simons forms
\begin{equation}
d_{a_1a_2...a_k}H^{a_1}\wedge H^{a_2}\wedge ...\wedge H^{a_{k-1}}\wedge
B^{a_k},  \label{ChernSimons}
\end{equation}
where $d_{a_1a_2...a_k}$ is completely antisymmetric and where $H^{a_i}$
stands for the three-form $dB^{a_i}$ with $B^{a_i}=(1/2)B_{\mu \nu
}^{a_i}dx^\mu \wedge dx^\nu $. One has $n=2+3(k-1)$ (so that (\ref
{ChernSimons}) is a $n$-form), with $k\geq 2$, and of course, $k\leq N$.
This completes the analysis of the $n\geq 5$ case.\ \ In the
language of \cite{GWeinberg}, Theorem 2 shows that  the
structural constraint expressing
that the gauge symmetry should be  unchanged under
renormalization is fulfilled (Theorem 2 remains true if one
starts with an action already modified by gauge-invariant interactions \cite
{HKS2}).

Let us turn now to the case $n=4$, which is more complicated because the
cohomological group $H_2(\delta |d)$ differs from zero \cite{HKS1}. To
analyse this case, it is convenient to introduce the auxiliary field $%
A^{a\mu }\sim \varepsilon ^{\mu \nu \rho \sigma }H_{\nu \rho \sigma }^a$ and
to replace $B_{\mu \nu }^a$ by its dual $\frac 12\varepsilon ^{\mu \nu \rho
\sigma }B_{\rho \sigma }^bg_{ab}$, which we shall still denote by $B_a^{\mu
\nu }$.\ \ This is permissible because auxiliary fields do not change the
dynamics and, as it has been shown in \cite{MH1,BBH2}, they do not change
the local cohomological groups $H(s|d)$ either. The starting point is thus
the action with auxiliary fields, which read explicitly,
\begin{equation}
S_0^{\prime }=\frac 12\int d^4x\ \left( A_\mu ^aA^{b\mu }g_{ab}+B_a^{\mu \nu
}F_{\mu \nu }^a\right)
\end{equation}
with
\begin{equation}
F_{\mu \nu }^a=\partial _\mu A_\nu ^a-\partial _\nu A_\mu ^a.
\end{equation}
In terms of the new variables, the BRST differential $s=\delta +\gamma $ of
the free theory reads
\begin{eqnarray}
\delta A_\mu ^a &=&0,\ \gamma A_\mu ^a=0, \\
\delta B_a^{\mu \nu } &=&0,\ \gamma B_a^{\mu \nu }=\varepsilon ^{\mu \nu
\rho \sigma }\ \partial _\rho C_{a\sigma }, \\
\delta C_{a\mu } &=&0,\ \gamma C_{a\mu }=\partial _\mu \eta _a, \\
\delta \eta _a &=&0,\ \gamma \eta _a=0, \\
\delta A_a^{*\mu } &=&g_{ab}A^{b\mu }+\partial _\nu B_a^{\mu \nu },\ \gamma
A_a^{*\mu }=0, \\
\delta B_{\mu \nu }^{*a} &=&\frac 12F_{\mu \nu }^a,\ \gamma B_{\mu \nu
}^{*a}=0, \\
\delta C^{*a\mu } &=&\varepsilon ^{\mu \nu \rho \sigma }\partial _\nu
B_{\rho \sigma }^{*a}\ ,\gamma C^{*a\mu }=0, \\
\delta \eta ^{*a} &=&\partial _\mu C^{*a\mu },\ \gamma \eta ^{*a}=0.
\end{eqnarray}
The solution of the master equation for the free theory with auxiliary
fields is given by
\begin{equation}
S=S_0^{\prime }-\int d^4x\left( \varepsilon ^{\mu \nu \rho \sigma }\partial
_\rho C_{a\sigma }B_{\mu \nu }^{*a}+\partial _\mu \eta _aC^{*a\mu }\right)
\end{equation}
and of course, $s\cdot =(S,\cdot )$.

Let us now construct the most general $s$-cocycle $a$ modulo $d$ at ghost
number zero. In four dimensions, the groups $H_k(\delta |d)$ vanish also for
$k>3$ \cite{BBH2}. Thus, the cocycle $a$ can be assumed to have the expansion
$%
a=a_0+a_1+a_2$, where $a_2$ is of the form $\mu ^a\eta _a$ with $\mu ^a$ an
element of $H_2(\delta |d)$ that depends on $A_\mu ^a$, the antifields and
their derivatives. The most general element in $H_2(\delta |d)$ is, up to
trivial terms, given by
\begin{equation}
\mu ^a=\left( C^{*b\mu }A_\mu ^c+\frac 12B_{\mu \nu }^{*b}B_{\rho \sigma
}^{*c}\varepsilon ^{\mu \nu \rho \sigma }\right) f_{\ bc}^a
\end{equation}
where $f_{\ bc}^a=-f_{\ cb}^a$. It is easy to verify that $\mu ^a$ defines a
non trivial element of $H_2(\delta |d)$, and that there are no others. This
last assertion follows from the isomorphism between $H_0^1(d|\delta )$ and $%
H_3^4(\delta |d)$ \cite{BBH2} and will be verified explicitly in \cite{HKS1}.

The next step is to calculate $a_1$. One has
\begin{equation}
\delta a_2=\gamma \left( B_{\mu \nu }^{*a}\varepsilon ^{\mu \nu \rho \sigma
}f_{\ ab}^cA_\rho ^bC_{c\sigma }\right) +\partial _\mu V^\mu
\end{equation}
for some $V^\mu $. This determines $a_1$ to be
\begin{equation}
a_1=-B_{\mu \nu }^{*a}\varepsilon ^{\mu \nu \rho \sigma }f_{\ ab}^cA_\rho
^bC_{c\sigma }
\end{equation}
up to irrelevant trivial terms. By an identical calculation, one then finds $%
\delta a_1=-\gamma a_0+\partial _\mu k^\mu $, with
\begin{equation}
a_0=-\frac 12B_a^{\mu \nu }f_{\ bc}^aA_\mu ^bA_\nu ^c.
\end{equation}
By construction, $a=\mu ^a\eta _a+a_1+a_0$ is an $s$-cocycle modulo $d$ from
which one cannot eliminate the antifields because the last term $\mu ^a\eta
_a$ is unremovable. With this choice of $a$, $S+g\int d^4x\ a$ solves the
master equation up to order $g$.

As recalled above \cite{BH1}, one can add to $S+g\int d^4x a$ terms of
order $g^2$ in such a way that the master equation holds up to that order
iff the antibracket $\left( S+g\int d^4x\ a,\ S+g\int d^4x\ a\right) $is $s$%
-exact. A direct calculation yields for this antibracket
\begin{equation}
g^2\int d^4x\ t_{\ bde}^a\varepsilon ^{\mu \nu \alpha \beta }\left( B_{\mu
\nu }^{*b}\eta _a-\frac 13A_\mu ^bC_{a\nu }\right) A_\alpha ^dA_\beta ^e
\label{constraint}
\end{equation}
with
\begin{equation}
t_{\ bde}^a=f_{\ bc}^af_{\ de}^c+f_{\ dc}^af_{\ eb}^c+f_{\ ec}^af_{\ bd}^c.
\end{equation}
One verifies easily that the integrand is BRST-closed modulo $d$. However,
it is BRST-exact modulo $d$ if and only if the constants $f_{\ bc}^a$
fulfill the Jacobi identity,
\begin{equation}
t_{\ bde}^a=0,
\end{equation}
i.e., define the structure constants of a Lie algebra.\ Indeed, \ (\ref
{constraint}) is trivial if and only if the coefficient of $\eta _a$ is the $%
\delta $-variation of an invariant term modulo the divergence of an
invariant current, and this is possible only if it is zero.

In this case, (\ref{constraint}) vanishes altogether and thus $S+g\int d^4x\
a$ is a solution of the master equation by itself, without need for further
terms of higher order in $g$. The antifield-independent piece of $S+g\int
d^4x\ a$ is the familiar action $S_0^{\prime }+g\int d^4x\ a_0$ obtained by
Freedman and Townsend \cite{FT} (in first-order form), and the solution $%
S+g\int d^4x\ a$ of the master equation is the corresponding solution
constructed by various authors \cite{1,2,3}. Note that the metric $g_{ab}$
need not be invariant under the adjoint action of the Lie algebra. Note also
that  one can of course  add to $a_0$  an arbitrary polynomial in $A_\mu ^a$
and its derivatives, which is the most general gauge-invariant term, but
such additional interactions does not affect the gauge transformations.\ \
This completes the construction of the most general deformation of the
action for a system of free two-forms in four dimensions. We emphasize again
that the Freedman-Townsend vertex truly deforms the gauge symmetry, even up
to redefinitions, because the antifields cannot be removed from $a$.

In this letter, the most general interactions that can be introduced for a
system of free two-form gauge fields have been constructed.\ \ It has been
shown in particular that the form of the gauge symmetries is quite rigid, in
contrast to what happens for a system of
free one-forms. Indeed, the interactions
cannot change the gauge transformations, except in four dimensions, where
the only exception is given by the Freedman-Townsend vertex.

The analysis can be extended to $p$-forms, where one can also show that the
abelian gauge transformations are quite rigid. Similarly, by using the same
cohomological techniques, gauge invariance of all the conserved currents,
except those associated with the global rotations of the two-forms among
themselves, can be established along the lines of \cite{BBH3}.

\section*{Acknowledgements}

Discussions with Bernard\ Knaepen and Christiane\ Schomblond are gratefully
acknowledged.
This work has been supported in part by research funds from the F.N.R.S. and
a research contract with the Commission of the European Communities.

\end{document}